\documentclass[twoside,12pt]{article}
\usepackage{epsfig}
\usepackage{amsmath,amsfonts}

\def\d{\textrm{d}}
\def\e{\textrm{e}}

\newcommand{\be}{\begin{equation}}
\newcommand{\ee}{\end{equation}}
\newcommand{\bea}{\begin{eqnarray}}
\newcommand{\eea}{\end{eqnarray}}

\begin{document}

\title{Statistical measure of complexity and  correlated behavior of Fermi systems}

\author{Ch.C. Moustakidis$^{1}$, V.P. Psonis$^{1}$, K.Ch. Chatzisavvas$^{1,2}$, C.P. Panos$^{1}$, and S.E. Massen$^{1}$ \\
$^{1}$Department of Theoretical Physics, Aristotle University of
Thessaloniki, \\ 54124 Thessaloniki, Greece \\
$^{2}$Informatics and Telecommunications Engineering Department,
\\ University of Western Macedonia, 50100 Kozani, Greece }

\maketitle

\begin{abstract}
We apply  the statistical measure of complexity, introduced by
L\'{o}pez-Ruiz, Mancini and Calbet (LMC), to uniform Fermi
systems. We investigate the connection between information and
complexity measures with the strongly correlated behavior of
various Fermi systems as nuclear matter, electron gas and liquid
helium. We examine the possibility that LMC complexity can serve
as an index quantifying correlations in the specific system and to
which extent could be related with experimental quantities.
Moreover, we concentrate on thermal effects on the complexity of
ideal Fermi systems. We find that complexity behaves, both at low
and high values of temperature, in a similar way as the specific
heat.
\vspace{0.3cm}

PACS number(s): 05.30.Fk, 89.70.Cf, 05.20.-y, 05.90.+m  \\


\end{abstract}

\section{Introduction}

Information theory, introduced by Shannon in 1948 to provide
answers for some fundamental aspects of communications
\cite{Shannon48}, celebrates its first massive application in a
quantum system, i.e. atoms, almost four decades later
\cite{Gadre84}. In his seminal work, Gadre stated that
\emph{information theory is a hidden treasure yet to be
discovered},  which is still true. His prediction has been
justified over the past twenty years by the extended applications
of information-theoretic methods
\cite{Shannon48,Onicescu66,Bialynicki75,Fisher25,Kullback59}, in
various quantum systems. Information-theoretical methods play an
important role, not just in the clarification of fundamental
concepts of quantum mechanics, but also provide a series of
results concerning the information content of systems, the
presence of interactions, correlation with experimental measured
quantities, extraction of universal relations etc
\cite{Gadre84,Ohya93,BAB95,Nagy96,Majer96,Panos97,Massen98,Ghosh84,Garba06,Frieden04,Moustakidis05,Patil07,Liu07,Luzanov07,Antolin08,Sagar08}.
Additionally, many complexity measures have been proposed as
indicators of  complex behavior found in different systems
scattered in a broad spectrum of fields
\cite{Lopez95,Catalan02,Sanudo09,Landsberg-98,Shiner-99,Crutch98,Crutch00,Binder00,Chatzisavvas05,Martin03,
Chatzisavvas07a,Chatzisavvas07b,Panos-09,Lopez05,Yamano04,Plastino96,Borgoo07,Montogomery08,Calbet-07,
Angulo08,Angulo08b,Sanudo08,Sanudo08b,Chatzi-09,Roza-09-a,Ruiz-09,Roza-09-b}.

The statistical measure of complexity $C_{LMC}$ introduced by
L\'{o}pez-Ruiz, Calbet and Mancini (LMC) \cite{Lopez95} identifies
the entropy or information stored in a system and its distance to
the equilibrium probability distribution as the two basic
ingredients giving the correct asymptotic properties of a
well-behaved measure of complexity. So far several complexity
measures have been proposed as indicators of complex behavior in
various systems, mostly coming from physics, computational
sciences e.t.c. The LMC measure of statistical complexity is an
easily calculable measure (compared with other definitions e.g.
Kolmogorov's one), defined in the form of the product $C_{\rm
LMC}=S D$, combining information $S$ and \emph{disequilibrium}
$D$. It has indeed the features and asymptotical properties that
one expects intuitively. It vanishes for the two extreme cases of
a perfect crystal (perfect order) and ideal gas (perfect
disorder). The initial definition of $C_{LMC}$ has been slightly
modified in a suitable way by Catalan \emph{et al}
\cite{Catalan02}, leading to the form $C=\e^S D$ applicable to
systems described by either discrete or continuous probability
distributions. In \cite{Catalan02} it was shown that the results
in both, discrete and continuous cases, are consistent: extreme
values of $C$ are observed for distributions characterized by a
peak superimposed onto a uniform sea. Moreover, $C$ should be
minimal, when the system reaches equipartition and the minimum
value of $C$ is attained for rectangular (uniform) density
probabilities giving the value $C=1$. Additionally, $C$ is not an
upper bounded function and can become infinitely large.

LMC complexity is referred to in the literature as shape
complexity, since it exhibits larger values for complicated
patterns of probabilities, as seen, in a first step, intuitively
and by inspecting the plots. The first investigation of $C_{LMC}$
in quantum many-body systems was carried out in atoms, for
continuous electron distributions \cite{Chatzisavvas07b} and
discrete ones \cite{Panos-09}. An alternative definition of
complexity is the SDL measure $\Gamma_{\alpha \beta}$
\cite{Shiner-99} (Shiner, Davison, Landsberg), defined and
calculated in an analogous way as the LMC one. It has been applied
in atoms as well, starting from \cite{Chatzisavvas07a}. Comments
on the validity of SDL and LMC measures are given in detail in
Section~4 of \cite{Chatzisavvas07b}, where it is stated that a
welcome property of a definition of complexity might be the
following: If one complicates the system by varying some of its
parameters, and this leads to an increase of the adopted measure
of complexity, then one could argue that this measure describes
the complexity of the system properly.

The modified version $C=\e^S D$, also known as shape complexity,
satisfies some additional and desirable features such as
positivity, invariance under translations, rescaling
transformations and replication. Also, another indication of its
internal consistency is the fact that the first two $q$-values of
the R\'{e}nyi entropy \cite{Renyi61} are the two defining elements
of shape complexity, i.e. $\e^S$ and $D$ \cite{Antolin09}. The
usefulness of the improved version has been shown in many fields
\cite{Binder00,Chatzisavvas05,Martin03,
Chatzisavvas07a,Chatzisavvas07b,Panos-09,Lopez05,Yamano04,Plastino96,Borgoo07,Montogomery08,Calbet-07,
Angulo08,Angulo08b,Sanudo08,Sanudo08b,Chatzi-09}. Moreover, the
specific measure is suitably tailored for quantum systems,
described by their very nature probabilistically via density
distributions in position and momentum spaces, which are necessary
for and enable a relatively easy calculation of $S$ and $D$,
entering the formulas $C_{\rm LMC}=S D$ or $C=\e^S D$.

The motivation of the present work,  in the spirit of the above
statements, is to extend our previous study  of uniform Fermi
systems \cite{Moustakidis05}, beyond information entropy, in order
to include the complexity measure proposed by L\'{o}pez-Ruiz et
al. \cite{Lopez95}, using probability distributions in momentum
space. In uniform systems the density $\rho=N/V$ is a constant and
the interaction of the particles is reflected to the momentum
distribution which deviates from the $theta$ function form of the
ideal Fermi-gas model. Our aim is to connect $C$, a measure based
on a probabilistic description and the shape of the corresponding
momentum distributions to the phenomenological parameters
introducing the inter-particle correlations and experimental data
(e.g. specific heat). It is important to examine how the
interaction affects the momentum distribution as well as the
complexity. An attempt is also made to relate the complexity $C$
with statistical quantities such as the temperature.

The complexity $C$ cannot be measured experimentally, but it is
possible, as we demonstrate here, to calculate it, starting from a
reasonable definition (LMC or SDL) and in our case, by employing
an information-theoretic method, developed in previous work
\cite{Moustakidis05}. Experiment can enter our work through the
experimentally measured momentum distribution $n(k)$. Momentum
distributions $n(k)$ can be assessed experimentally (e.g. in
nuclear matter and liquid helium) via deep inelastic scattering at
a large momentum transfer. The extraction of $n(k)$ from the
measured scattering intensity is influenced by the limitations
imposed by the experimental resolution and the final-state
interaction. Hence, the most accurate information on $n(k)$
available to date is likely to be one obtained through accurate
theoretical calculations, aided by experimental input.

In the present work,  the quantum systems under examination are
nuclear matter, electron gas and liquid $^3$He. The inter-particle
interactions of these systems differ in general by many orders of
magnitude in strength and range. The corresponding potentials,
scaled under suitable energy and length measures for the different
systems, i.e. Fermi energy and inverse Fermi momentum, still
differ by orders of magnitude. The $^3$He system is the most
strongly interacting one at short distances with an
almost-hard-core interaction, while electron gas is the most
weakly interacting. Nuclear matter lies somewhere in between.
Helium and nuclear potentials have relatively weak attractive
tails. The electronic potential is quite different. It has a weak
core (compared with $^3$He and nuclear matter) but its  rate of
decrease for large $r$ is slow. Thus, at large distances, the
electronic potential is stronger than the other two.


Furthermore, the density under investigation affects the
assessment of the effect of strong versus weak interaction.
Characteristic is the example of the electron gas, which is
distinguished from other systems by the long range nature of the
Coulomb interaction. As a result, strong coupling prevails in the
limit of low density for electron gas, whereas helium and nuclear
systems become more strongly interacting for higher density
regions. In all cases the strength of the interaction may be
gauged by the depletion of the Fermi sea. Quantitatively, this can
be assessed as the deviation of $Z_{F}$ from unity, where $Z_{F}$
is the discontinuity gap of the momentum distribution $n(k)$ at
$k=k_{F}$, in a uniform Fermi system. The problem of discontinuity
of the momentum distribution is examined for Fermi-liquids by
Migdal \cite{Migdal-67}. It is shown that the discontinuity in the
momentum distribution at $k=k_{F}$ is an inherent consequence of
an arbitrary interaction between particles in an infinite system.

The paper is organized as follows. In Section I we present the
method leading to the expressions of momentum distribution,
information content and complexity measures in finite Fermi
systems. Applications of that expression to nuclear matter,
electron gas, and liquid $^3$He are made in the three subsections
of Section II. In the same subsections numerical results are also
reported and discussed. In Section III the study of the influence
of thermal effects on the complexity  is made. Finally, the
concluding remarks and a summary of the present work are given in
Section IV.

\section{Momentum distribution, information entropy and complexity}
The one-body density matrix is the key quantity for the
description of the momentum distribution, both in infinite and
finite quantum systems. It is defined as
\begin{equation}
    \rho({\bf r}_1,{\bf r}_1')= \int \Psi^{*}({\bf r}_1,{\bf r}_2,
    \ldots,{\bf r}_N) \, \Psi({\bf r}_1',{\bf r}_2,\ldots,{\bf r}_N)\,
    \d{\bf r}_2 \ldots \d{\bf r}_N.
\end{equation}
The diagonal elements $\rho({\bf r_1},{\bf r_1})$ of the density
matrix yield the local density distribution, which is just a
constant $\rho$ in the case of a uniform infinite system.
Homogeneity and isotropy of the system require that $\rho({\bf
r}_1,{\bf r}_1')= \rho(\mid {\bf r}_1-{\bf r}_1'\mid)\equiv
\rho(r)$.

The momentum distribution for fermions is defined by
\begin{equation}
n(k)=\nu^{-1} \int \rho(r) \, \e^{i {\bf kr}} \, \d{\bf r},
\end{equation}
where $\nu$ is the single-particle level degeneracy
\begin{equation*}
\nu = \left\{
\begin{array}{rl}
2 & \text{for electron gas and liquid $^3$He}\\
4 & \text{for nuclear matter.}
\end{array} \right.
\end{equation*}
The normalized momentum distribution,
\begin{align*}
    \int n(k) \, \d{\bf k}=1,
\end{align*}
is given by the relation
\begin{equation}
n(k)=\frac{1}{V_k}\tilde{n}(k)=\frac{1}{V_k}\left\{
\begin{array}{cc}
    \tilde{n}_-(k)   , &  \mbox{$k<k_F$ } \\
     \tilde{n}_+(k)      , &  \mbox{$k>k_F$,}
                              \end{array}
                       \right.
\label{Cor-MD}
\end{equation}
where $V_k=\frac{4}{3}\pi k_F^3$. The Fermi wave number $k_F$ is
related with the constant density $\rho=N \rho_0=3/(4\pi r_0^3)$
as follows
\begin{equation}
    k_F=\left( \frac{6 \pi^2 \rho}{\nu} \right)^{1/3}= \
    \left(\frac{9 \pi}{2 \nu}\frac{1}{r_0^3} \right)^{1/3}. \label{kfermi}
\end{equation}
In the case of an ideal Fermi gas the momentum distribution has
the form
\begin{equation}
    n_0(k)=\frac{1}{V_k}\, \theta(k_F-k). \label{MD-IG}
\end{equation}
The information entropy in momentum space is given by the relation
\begin{equation}
    S_k=-\int n(k) \ln{n(k)} \, \d{\bf k} . \label{IE-Sk-1}
\end{equation}
So, for an ideal Fermi gas, using Eq. (\ref{MD-IG}), $S_{k}$
becomes
\begin{equation}
S_k=S_0=\ln V_k=\ln \left(\frac{6 \pi^2}{\nu}\frac{1}{r_0^3}
\right). \label{IE-Sk-2}
\end{equation}

For correlated Fermi systems, the information entropy  in momentum
space, can be found from Eq. (\ref{IE-Sk-1}) if we replace $n(k)$
from Eq. (\ref{Cor-MD}). $S_k$ is written now \cite{Moustakidis05}
\begin{equation}
S_k=\ln V_k-\frac{4 \pi}{V_k}\left( \int_{0}^{k_F^{-}} k^2
\tilde{n}_-(k)\ln\tilde{n}_-(k)  dk+ \int_{k_F^{+}}^{\infty} k^2
\tilde{n}_+(k) \ln\tilde{n}_+(k)  dk \right). \label{Cor-Sk-1}
\end{equation}
The correlated entropy $S_k$ has the form
\begin{equation}
S_k=S_0+S_{\rm cor}, \label{Cor-S-1}
\end{equation}
where $S_0$ is the uncorrelated entropy given by
Eq.~(\ref{IE-Sk-2}) and $S_{\rm cor}$ is the contribution of the
particles correlations to the entropy. That contribution can be
found from the expression
\begin{equation}
S_{\rm cor}=-3\left( \int_{0}^{1^{-}} x^2
\tilde{n}_-(x)\ln\tilde{n}_-(x) dx+ \int_{1^{+}}^{\infty} x^2
\tilde{n}_+(x)\ln\tilde{n}_+(x) dx \right), \label{S-cor}
\end{equation}
where $x=k/k_F$.

The disequilibrium $D_k$ (or information energy, defined by
Onicescu \cite{Onicescu66}), in momentum space  is defined as
\begin{equation}
D_k=\int n^2(k) d{\bf k} . \label{D-1}
\end{equation}
and for an ideal Fermi gas, using Eq. (\ref{MD-IG}), becomes
\begin{equation}
D_k=D_0=\frac{1}{V_k}. \label{D-ideal}
\end{equation}
In the case of correlated Fermi systems, $D_k$ is written as
\begin{equation}
D_k=\frac{1}{V_k} \frac{4\pi}{V_k}\left( \int_{0}^{k_F^{-}} k^2
\tilde{n}^2_-(k) dk+ \int_{k_F^{+}}^{\infty} k^2
\tilde{n}^2_+(k)dk \right). \label{D-Cor-1}
\end{equation}
The correlated disequilibrium $D_k$ has the form
\begin{equation}
D_k=D_0 \, D_{\rm cor}, \label{D-Cor-2}
\end{equation}
where $D_0$ is  given in  Eq. (\ref{D-ideal}) and $D_{cor}$  can
be found from the expression
\begin{equation}
D_{cor}=3\left( \int_{0}^{1^{-}} x^2 {n}^2_-(x)dx+
\int_{1^{+}}^{\infty} x^2 {n}^2_+(x) dx \right). \label{D-cor-3}
\end{equation}
%


The LMC statistical measure of complexity $C_{LMC}$, in momentum
space, is defined as \cite{Lopez95}
\begin{equation*}
    C_{\rm LMC}= S_{k} D_{k},
\end{equation*}
where $S_{k}$ is the Shannon information entropy, while $D_{k}$ is
the disequilibrium of the system under investigation (in momentum
space).

The modified version of the complexity, proposed by Catalan et.
al. \cite{Catalan02}, in momentum space, is defined as
\begin{equation}
    C=H_k D_k, \label{C-1}
\end{equation}
where  $H_k$ represents the information content of the system
defined as
\begin{equation}
    H_k=\e^{S_k}, \label{H-1}
\end{equation}
and ensures positivity of information under any circumstances.

It is easy to show that
\begin{equation}
    C=C_0 \,  C_{\rm cor}=\e^{S_{\rm cor}} D_{\rm cor} , \qquad C_0=\e^{S_0} D_0 =1. \label{C-2}
\end{equation}
The physical meaning of Eq.(\ref{C-2}) is clear. In the case of an
ideal Fermi gas (see Eq.~(\ref{MD-IG})) $C$ is minimal with value
$C_0=1$ (see also \cite{Catalan02}). Moreover as pointed out in
Ref.~\cite{Catalan02}, $C$ is not an upper bounded function and
can therefore become infinitely large. From the above analysis it
is clear that complexity $C$ is an accounter of correlations in an
infinite Fermi system. So, the next step is to try to find the
connection between $C$ and the correlation parameters of the
systems. The correlations invoke diffusion of the momentum
distribution and we expect this effect to be reflected on the
values of $C$.

Here, we apply the low order approximation of the momentum
distribution in the case of the nuclear matter
\cite{Gaudin71,Dalri82,Flyn84}. For liquid $^3$He and electron gas
we use the most updated calculations for the momentum
distribution, that is the results of Moroni et al. \cite{Moroni97}
and P. Gori-Giorgi et al. \cite{Paola-02}, respectively. Also, we
would like to stress out the fact that our main goal is the
accurate calculation of the correlated part of information and
complexity measures, based in reliable data, and not the detailed
analysis of the momentum distribution itself.

\subsection{Nuclear matter}

The model we study is based on the Jastrow ansatz for the ground
state wave function of nuclear matter
\begin{equation}
\Psi({\bf r}_1,{\bf r}_2,...,{\bf r}_N)= \prod_{1 \leq i \leq  j
\leq N} f(r_{ij}) \, \Phi({\bf r}_1,{\bf r}_2,...,{\bf r}_N),
\label{Jastrow-1}
\end{equation}
where $r_{ij}=\mid {\bf r}_i-{\bf r}_j \mid$, $\Phi$ is a Slater
determinant (here, of plane waves with appropriate spin-isospin
factors, filling the Fermi sea) and $f(r)$ is  a state-independent
two-body correlation function. Thus, the correlation function is
taken to be of the Jastrow type \cite{Jastrow55}
\begin{equation}
f(r)=1-\exp[-\beta^2 r^2], \label{beta-Gaus}
\end{equation}
where $\beta$ is the correlation parameter. A cluster expansion
for the one-body density matrix $\rho({\bf r}_1,{\bf r}_1')$ has
been derived by Gaudin, Gillespie and Ripka
\cite{Gaudin71,Dalri82,Flyn84} for the Jastrow trial function
(\ref{Jastrow-1}).

In the low order approximation the momentum distribution is
constructed as \cite{Flyn84}
\begin{equation}
n_{\rm LOA}(k)=\theta(k_F-k) \left[ 1-k_{\rm dir}+Y(k,8) \right]+
8 \left[ k_{\rm dir}Y(k,2)-[Y(k,4)]^2 \right], \label{mn-mom-1}
\end{equation}
where
\begin{equation}
    c_{\mu}^{-1}Y(k,\mu)=
    \frac{\e^{-\tilde{k}_{+}^{2}}-\e^{-\tilde{k}_{-}^{2}}}{2\tilde{k}}
    +\int_{0}^{\tilde{k}_{+}} \e^{-y^2} \, \d y +
    {\rm sgn}(\tilde{k}_{-}) \int_{0}^{\mid \tilde{k}_{-} \mid} \e^{-y^2} \d y,
\end{equation}
and
\begin{equation}
    c_{\mu}=\frac{1}{8\sqrt{\pi}}\left(\frac{\mu}{2}\right)^{3/2},
    \quad \tilde{k}=\frac{k}{\beta \sqrt{\mu}} , \quad
    \tilde{k}_{\pm}=\frac{k_{F}\pm k}{\beta \sqrt{\mu}}, \quad
    \mu=2,4,8.
\end{equation}
while ${\rm sgn}(x)=x/\mid x \mid$. The dimensionless Jastrow
wound parameter $k_{\rm dir}$ can serve as a rough measure of
correlations and the rate of convergence of the cluster expansion
is defined as
\begin{equation}
    k_{\rm dir}=\rho \int [f(r)-1]^2 \, \d {\bf r}, \label{eq-kdir}
\end{equation}
The normalization condition for the momentum distribution is
\begin{equation}
\int_{0}^{\infty} n_{\rm LOA}(k)k^2 \d k=\frac{1}{3}\, k_{F}^{3}.
\end{equation}
>From  Eq.~(\ref{eq-kdir}) we obtain the following relation between
the wound parameter $k_{\rm dir}$ and the correlation parameter
$\beta$
\begin{equation}
k_{\rm dir}=\frac{1}{3 \sqrt{2\pi}}\left(\frac{k_F}{\beta}
\right)^3.
\end{equation}
It is clear that large values of  $k_{\rm dir}$ imply strong
correlations and poor convergence of the  cluster expansion. In
the numerical calculations the correlation parameter $\beta$ is in
the interval: $1.01 \leq \beta \leq  2.482 $. That range
corresponds to $ 0.3 \geq k_{\rm dir}  \geq 0.02 $ and is
reasonable in the case of nuclear matter \cite{Flyn84}.

The calculated values of $S_{\rm cor}$, $D_{\rm cor}$ and $C$ for
nuclear matter versus wound parameter $k_{\rm dir}$ are displayed
in Fig.~1. $S_{\rm cor}$ and $C$ increase with $k_{\rm dir}$,
while $D_{\rm cor}$ decreases. We fitted the numerical values of
the above quantities, with simple functions of $k_{\rm dir}$ and
we find respectively the following formulae
\begin{equation}
    S_{\rm cor}=\alpha k_{\rm dir}^{\beta}, \qquad \alpha=2.0586, \quad
    \beta=0.6365. \label{Scor-fit-1}
\end{equation}
\begin{equation}
D_{\rm cor}=1+\alpha k_{\rm dir}^{\beta}, \qquad \alpha=-0.9009,
\quad \beta=0.8325. \label{Dcor-fit-1}
\end{equation}
\begin{equation}
C=1+\alpha k_{\rm dir}^{\beta} \e^{\gamma k_{\rm dir}}, \qquad
\alpha=3.1760, \quad \beta=0.8257, \quad \gamma=-1.6176.
\label{Ccor-fit-1}
\end{equation}
The values of the parameters $\alpha$, $\beta$ and  $\gamma$, for
each case, have been selected by a least squares fit (LSF) method.

Another characteristic quantity, used as a measure of the strength
of correlations of the uniform Fermi systems, is the
discontinuity, $Z_F$, of the momentum distribution at $k/k_F=1$.
It is defined as
\begin{equation*}
    Z_F=n(1^-)-n(1^+)
\end{equation*}

The behavior of momentum distribution, as a function of $k/k_F$
for various values of the wound parameter $k_{\rm dir}$ is
indicated in Fig.~2(a). The discontinuity $Z_F$ is also displayed
in each case. For ideal Fermi systems $Z_F=1$, while for
interacting ones $Z_F<1$. In the limit of very strong interaction
$Z_F=0$, there is no discontinuity on the momentum distribution of
the system. The quantity ($1-Z_F$) measures the ability of
correlations to deplete the Fermi sea by exciting particles from
states below it (hole states) to states above it (particle states)
\cite{Flyn84}.

The dependence of $S_{\rm cor}$, $D_{\rm cor}$ and  $C$ on the
quantity $(1-Z_F)$ is shown in Fig.~3. It is seen that $S_{\rm
cor}$ and $C$ are increasing functions of $(1-Z_F)$, while $D_{\rm
cor}$ is a decreasing one, as a direct consequence of the
dependence of the above quantities, on the correlation parameter
$k_{\rm dir}$. That dependence can be reproduced very well by
simple expressions as in Eqs.~(\ref{Scor-fit-1}),
(\ref{Dcor-fit-1}) and (\ref{Ccor-fit-1}) replacing  $k_{\rm dir}$
by $(1-Z_F)$. For example the expression
\begin{equation}
C(Z_F)=1+\alpha (1-Z_F)^{\beta} \, \e^{\gamma (1-Z_F)}
\label{C-Zf-NM}
\end{equation}
with $\alpha=3.6227$, $\beta=0.8024$ and $\gamma=-1.9750$
reproduces the numerical values of $C$ very well.

>From the above analysis we can conclude that LMC complexity $C$
can be employed as a measure of the strength of correlations in
the same way the wound and the discontinuity parameters are used.
An explanation of the above behavior of $C$ is the following: The
effect of nucleon correlations is the departure from the step
function form of the momentum distribution (ideal Fermi gas) to
the one with long tail behavior for $k>k_F$. The diffusion of the
distribution leads to a decrease of the order of the system (the
disequilibrium $D_k$ decreases and the information entropy $S_k$
increases accordingly). In total, the contribution of $S_k$ in $C$
dominates over the contribution of $D_k$ and thus the complexity
increases with the correlations (at least in the region under
consideration).

\subsection{Electron gas}

We consider as electron gas, a system of fermions interacting via
a Coulomb potential. The electron gas is a model of conduction
electrons in a metal, where the periodic positive potential due to
the ions is replaced by a uniform charge distribution. The density
of the uniform electron gas (Jellium) is $\rho=3/(4\pi r_o^3)$ and
the momentum distribution is $n(x,r_s)$, a function of both
$x=k/k_F$ and $r_s=r_o/a_B$ (where $a_B=\hbar^2/me^2$ is the Bohr
radius). In the Fermi-liquid regime, the momentum distribution of
the unpolarized uniform electron gas $n(x,r_s)$ is constructed
with the help of the convex Kulik function $G(\chi)$
\cite{Paola-02}.

Discontinuity $Z_F(r_s)$ is unit for $r_s =0$ and it is a
decreasing function of interaction strength $r_s$, as $r_s$
increases. The discontinuity gap of the momentum distribution
$n(k)$ at the Fermi surface narrows as the density decreases, a
clear indication that the system becomes more strongly coupled.
That behavior is due to the fact that the screening of the
long-range Coulomb interaction between the electrons becomes less
effective at lower density. Nuclear matter and atomic $^3$He
exhibit an inverse behavior, where the basic interactions are of
short range and $Z_{F}$ decreases as the density increases. At
large $r_s$, electrons form a Wigner crystal with a smooth
$n(x,r_s)$. Interaction strength $r_s \ll 1$ is the
weak-correlation limit and $r_s \gg 1$ is the strong-correlation
limit, respectively. For intermediate values of $r_s$, a non-Fermi
liquid regime may exist with $Z_F=0$ \cite{Paola-02}.

We examine the dependence of $S_{\rm cor}$, $D_{\rm cor}$ and $C$
for the electron gas on the correlation parameter $r_s$, (or
$\rho=3/(4\pi r_o^3)$) and  the discontinuity parameter ($1-Z_F$).
This dependence is displayed in Fig.~4. It is seen that, as in the
case of nuclear matter, $S_{cor}$ depends on those quantities
through a two parameter expression of the form
\begin{equation}
S_{\rm cor}(r_s)=\alpha r_s^{\beta},  \quad \alpha=0.1312, \qquad
\beta=0.8648. \label{Scor-ro-EG}
\end{equation}
The disequilibrium $D_{\rm cor}$ takes the form
\begin{equation}
D_{\rm cor}=1-\alpha r_s, \qquad  \alpha=0.0463, \label{Dcor-EG-1}
\end{equation}
and the complexity $C$ behaves as
\begin{equation}
C=1+\alpha r_{s}^{\beta} \e^{\gamma r_{s}}, \qquad \alpha=0.0700,
\quad \beta=1.4144, \quad \gamma=-0.1525. \label{Ccor-EG-1}
\end{equation}
The most distinctive feature, in the case of electron gas, is the
occurrence of a maximum of $C$ for high values of the correlations
parameter $r_s$ in contrast to the case of nuclear matter and
liquid helium (see bellow), where $C$ is a monotonic increasing
function of correlations. For high values of $r_s$, the
competition between $D_{\rm cor}$ and $\e^{S_{\rm cor}}$ (see
Eq.~(\ref{C-2})) leads to the dominance of the trend of $D_{\rm
cor}$. More precisely the slope of $C$ is given by
\begin{equation}
\frac{dC}{dr_s}=C\left(\frac{d \ln D_{\rm
cor}}{dr_s}+\frac{dS_{\rm cor}}{dr_s} \right). \label{dC-1}
\end{equation}
Thus, according to Eq.~(\ref{dC-1}) the sign of the slope of $C$
depends on the sum of the terms $\frac{d \ln D_{\rm cor}}{dr_s}$
and $\frac{dS_{\rm cor}}{dr_s}$ (which are always negative and
positive respectively when $r_s$ increases). It is easy to show by
applying Eqs.~(\ref{Scor-ro-EG}) and (\ref{Dcor-EG-1}) (or
equivalently, but with less accuracy from Eq.~(\ref{Ccor-EG-1}))
that $C$ attains a maximum value $C_{\rm max}\simeq 1.4052$ at
$r_s \simeq 9.589$.

The above feature is well reflected also on the $1-Z_F$ dependence
of $C$ as exhibited in Fig.~3. (the trend of $S_{\rm cor}$ and
$D_{\rm cor}$ is also shown).

\subsubsection{Momentum distribution and complexity for the Wigner Crystal}
In the low-density limit, $r_s \rightarrow \infty$, the electron
gas undergoes Wigner crystallization. The momentum distribution of
the localized electron is of harmonic-oscillator type and has the
form
\begin{equation}
    n(k,r_s \rightarrow \infty)=\left(\frac{1}{z} \frac{1}{\pi k_F^2 }\right)^{3/2} \,
    \e^{-\frac{k^2}{k_F^2}\frac{1}{z}},\qquad 4\pi\int_0^{\infty}k^2
    n(k,r_s \rightarrow \infty) \d k=1, \label{Wign-1}
\end{equation}
where $z=\omega/k_F^2=0.24r_s^{1/2}$ \cite{Paola-02}. It is easy
to prove that after some algebra
\begin{equation}
    S_{\rm cor}=\frac{3}{2}+\ln{\frac{3 \pi^{1/2}}{4}} +\frac{3}{2} \, \ln{z}.
\label{S-cor-wig}
\end{equation}
Additionally, $D_{\rm cor}$ can be found also from the expression
\begin{equation}
D_{\rm cor}=\frac{2^{1/2}}{3 \pi^{1/2}} z^{-3/2}.
\label{D-cor-wig}
\end{equation}
>From Eqs.~(\ref{S-cor-wig}), (\ref{D-cor-wig}) we find that
\begin{equation}
C= \e^{S_{\rm cor}} D_{\rm cor}
=\left(\frac{\e}{2}\right)^{3/2}\simeq 1.5845. \label{C-cor-wig-2}
\end{equation}
So, in the low-density limit (very strong correlations) the
complexity is independent of the correlation parameter $r_s$. As
we show below the above case is similar to that of an ideal Fermi
gas at high values of temperature.

\subsection{Liquid  $^3$He}

The interaction potential for liquid $^3$He is very strong at
small distances and its core repulsion is very hard (but not
infinite). Consequently there is a Fermi surface discontinuity of
roughly $Z_F\sim 0.3$. This small value supports the view that
$^3$He is the most strongly interacting Fermi system of the three
systems considered here. The momentum distribution has been
calculated using diffusion Monte Carlo simulations with the use of
trial functions optimized via the Euler-Monte Carlo method
\cite{Moroni97}.

We examine the dependence of $S_{\rm cor}$, $D_{\rm cor}$ and $C$
on the density $\rho=3/(4\pi r_o^3)$ and  the discontinuity
parameter $(1-Z_F)$, and we present our results in Fig.~5. The
dependence of $S_{\rm cor}$ on those parameters is described
through the following simple two parameter formula (as in the
cases of electron gas and nuclear matter)
\begin{equation} S_{cor}(\rho_0)=\alpha
\rho_0^{\beta}, \qquad \rho_0=100\rho, \label{Scor-rho-HL}
\end{equation}
with
\[\alpha=2.2736, \qquad \beta=1.4757, \]
while the  disequilibrium $D_{\rm cor}$ is described by the
function
\begin{equation}
D_{\rm cor}=\frac{\alpha}{1+\e^{(\rho_0-\beta)/\gamma}}+\delta,
\qquad \alpha=0.1321, \quad \beta=1.6288, \quad \gamma=0.1031,
\quad \delta=0.4062. \label{Dcor-LH-1}
\end{equation}
The complexity $C$ behaves as
\begin{equation}
    C=1+\alpha \rho_0^{\beta} \, \e^{\gamma \rho_0}, \qquad
    \alpha=0.0166, \quad \beta=-3.1849, \quad \gamma=5.8556.
\label{Ccor-LH-1}
\end{equation}
The dependence of $S_{\rm cor}$, $D_{\rm cor}$ and  $C$ on the
quantity $(1-Z_F)$ is seen in Fig.~3. In order to be able to
compare the results of the various systems, in the case of liquid
$^3$He the values of $S_{\rm cor}$ have been divided by $10$ and
the values of $C$ by $100$. The most distinctive feature of the
above analysis, in the various systems, is the different behavior
exhibited by $S_{\rm cor}$, $D_{\rm cor}$ and $C$ as  function of
$(1-Z_F)$. For the same values of $(1-Z_F)$ both the values and
the trend of these quantities are different in those  systems.


\section{Thermal effects on complexity in electron gas}
At temperature $T=0$, the electrons of the electron gas occupy all
the lower available states up to a highest one, namely the Fermi
level. As the temperature increases the electrons of the gas tend
to become excited and occupy states of energy of order $kT$ higher
than the Fermi energy. In general the occupation number of the
electron gas $n(\epsilon)$, is given by the Fermi-Dirac formula
\begin{equation}
    n(\epsilon)=\frac{1}{\exp\left[\frac{1}{k_B T}\,(\epsilon-\mu)
    \right] +1}, \label{Fermi-Dirac}
\end{equation}
where $\epsilon=\frac{p^2}{2m}$ ($p=\hbar k$) is the energy of the
electrons, $k_B$ is the Boltzmann's constant and $\mu$ is the
chemical potential. For $T=0$, the chemical potential of a gas
coincides with the Fermi energy $\epsilon_F$, which is by
definition the energy of the highest single-particle level
occupied at $T=0$. i.e.
\begin{equation}
    \epsilon_F=\frac{\hbar^2}{2m}\, \left(3\pi^2 \rho \right)^{2/3},
\end{equation}
while the Fermi temperature is defined via the  relation
\begin{equation}
    \epsilon_F=k_B T_F.
\end{equation}

We will examine how information and complexity measures considered
in the present study are affected by temperature, when it starts
to increase above zero. The limits of low temperature (quantum
regime) and high temperature (classical regime) are studied
separately in the following subsections.

\subsection{Quantum regime ($T\ll T_F$) }

With the term low energy we refer to the limit $T\ll T_F$, since
there is only a single characteristic value of temperature, the
Fermi temperature. In a first approximation, the chemical
potential for that limit, is \cite{Goodstein,Mandl,Huang}
\begin{equation}
    \mu=\epsilon_F\, \left[1-\frac{\pi^2}{12}\left(\frac{T}{T_F}\right)^2
\right],
\end{equation}
and so Eq. (\ref{Fermi-Dirac}) becomes
\begin{equation}
    n(x)=\frac{1}{\exp\left[\frac{1}{\xi}\,
    \left(x^2-1+\frac{\pi^2}{12}\, \xi^2\right)\right]+1}, \label{nx-1}
\end{equation}
where $x=(\epsilon/\epsilon_F)^{1/2}=k/k_F$, $\xi=T/T_F\ll 1$ and
$\int_{0}^{\infty} x^2n(x) \, \d x=1/3$.

Following the same procedure as in Section II, the information
entropy $S_k$ of the electron gas at temperature $T\ll T_F$ is
written
\begin{equation}
S_k=S_0+S_{\rm thermal},
\end{equation}
where $S_0$ is given by Eq.(\ref{IE-Sk-2}) and
\begin{equation}
S_{\rm thermal}=-3 \int_{0}^{\infty} x^2 n(x) \ln{n(x)} \, \d x.
\label{S-therm}
\end{equation}
In a similar way we find that $D_k$ is written as
\begin{equation}
D_k=D_0 D_{\rm thermal}, \label{D-Therm-2}
\end{equation}
where $D_0$ is  given in  Eq. (\ref{D-ideal}) and $D_{\rm
thermal}$ can be calculated also using the expression
\begin{equation}
    D_{\rm thermal}=3 \int_{0}^{\infty} x^2 n^2(x) \, \d x .\label{D-therm-3}
\end{equation}
Finally, it is easy to show that
\begin{equation}
C=C_0  C_{cor}=\e^{S_{\rm thermal}} \, D_{\rm thermal}  , \qquad
C_0=1. \label{C-therm-1}
\end{equation}

It is worthwhile to notice that the correlations between the Fermi
particles invoke a discontinuity to the momentum distribution at
$k=k_F$ (see Fig.~2(a)), while the thermal effect causes just a
slight deviation from the sharp step function form at $T=0$. This
is shown in Fig.~2(b), where the momentum distribution for an
ideal electron gas, at various values of $T/T_F$ has been plotted
versus $k/k_F$. The origin of the two effects (correlations and
temperature) is different and it is seen that they influence in a
different way the momentum distribution. So, it may be appropriate
to study qualitatively and also quantitatively the above effects
on the various information measures and complexity.

The calculated values of $S_{\rm thermal}$ for various values of
the temperature in the low energy limit ($T\ll T_F$) are shown in
Fig.~6. It is seen that $S_{\rm thermal}$ is an increasing linear
function of the temperature. The linear equation
\begin{equation}
S_{\rm thermal}=\alpha \, \frac{T}{T_F}, \qquad \alpha=2.5466
\label{IE-FG}
\end{equation}
reproduces very well the calculated values of $S_{thermal}$. That
expression of the information entropy is similar to the expression
which gives the thermodynamical entropy, $S_{TE}$, for $T\ll T_F$.
$S_{TE}$ in the low temperature limit has the form
\cite{Goodstein,Huang}
\begin{equation}
S_{TE}=\frac{\pi^2}{2}\, Nk_B \, \frac{T}{T_F}. \label{TE-FG}
\end{equation}
>From  Eqs.~(\ref{IE-FG}) and (\ref{TE-FG}), we can find a relation
between the two entropies in the limit $T\ll T_F$. The
corresponding relation is of the form
\begin{equation}
S_{\rm thermal}=\frac{2 \alpha}{\pi^2} \, \frac{S_{TE}}{N k_B}.
\label{IE-TE-FG-1}
\end{equation}
The calculated values of $D_{thermal}$, displayed in Fig.~6, are
well reproduced by the formula
\begin{equation}
D_{\rm thermal}=\e^{-1.647 \left(\frac{T}{T_F}\right)}.
\label{D-therm-low}
\end{equation}

Finally, the calculated values of  $C$ as a function of  $T/T_F$
are shown  in Fig.~6. $C$ is an increasing function of $T/T_F$ and
exhibits a linear  trend which is fairly reproduced by the
relation
\begin{equation}
C=1+0.962 \,\frac{T}{T_F}. \label{Ck-ther-fit}
\end{equation}

\subsection{Classical regime ( $T \gg T_F $)}

In the classical case (where the density is low and/or the
temperature is high and it is assumed that $n(k) \ll 1$) relations
can also be established between the information measures and
complexity with temperature.  In that case the momentum
distribution has the gaussian form \cite{Mandl}
\begin{equation}
n(k)=\left(\frac{a}{\pi}\right)^{3/2} \e^{-ak^2},\qquad
a=\frac{\hbar^2}{2 m k_B T}, \label{MD-idealgas}
\end{equation}
and is normalized as $\int n(k) \, \d {\bf k}=1$. The above
expression is also written as
\begin{equation}
    n(k)=\left(\frac{1}{\xi} \, \frac{1}{\pi k_F^2} \right)^{3/2}
    \e^{-\frac{k^2}{k_F^2}\frac{1}{\xi}}, \qquad \xi=T/T_F. \label{MD-idgas-2}
\end{equation}
>From Eqs. (\ref{S-therm}) and  (\ref{MD-idgas-2}), the following
relation, connecting $S_{\rm thermal}$ with the temperature, can
be found \cite{Moustakidis05}
\begin{equation}
S_{\rm thermal}=\frac{3}{2}+\ln{\frac{3 \pi^{1/2}}{4}}
+\frac{3}{2}\, \ln{\frac{T}{T_F}}. \label{Stherm-T}
\end{equation}
Additionally, we can also obtain $D_{\rm thermal}$ from the
expression
\begin{equation}
    D_{\rm thermal}=\frac{2^{1/2}}{3 \pi^{1/2}}\,
    \left(\frac{T}{T_F}\right)^{-3/2}. \label{D-Therm-3}
\end{equation}
>From Eqs.~(\ref{Stherm-T}), (\ref{D-Therm-3}) we find that
\begin{equation}
    C=C_{\rm thermal}= \e^{S_{\rm thermal}} \, D_{\rm thermal}
   =\left(\frac{\e}{2}\right)^{3/2}\simeq 1.5845.
\label{C-therm-2}
\end{equation}

The physical meaning of Eq.~(\ref{C-therm-2}) is very clear. For
high values of temperature ($T \gg T_F$), where the momentum
distribution of the gas is described by a Gaussian function, the
complexity $C$ is independent of $T$ and takes a constant value.
Our finding show that the case of an ideal Fermi gas at $T \gg
T_F$, is in contrast to the case of correlated Fermi gas at $T=0$,
where, in general, $C$ is not an upper bounded function. Moreover,
the complexity has different trend in the quantum mechanical limit
($T \rightarrow 0$) compared to the classical limit ($T
\rightarrow \infty$). In the classical limit, $C$ is not affected
by the temperature variation and is a constant of the system.
However, for low values of $T$, complexity exhibits strong
temperature dependence.

In Fig.~6 we plot the complexity versus $\xi=T/T_F$ both for low
and high values of temperature. Actually, we calculate the
chemical potential for each value of $\xi$ and consequently, we
know exactly the momentum dependence of the occupation numbers
given in Eq.~(\ref{Fermi-Dirac}). The results confirm the
numerical approximation of $C$ given in Eq.~(\ref{Ck-ther-fit})
for low values of $\xi$ as well as the analytical prediction of
Eq.~(\ref{C-therm-2}) for high values of $\xi$.

It is worthwhile to notice that the temperature dependence of $C$
is similar to that of the specific heat $C_V$ in an ideal Fermi
gas \cite{Huang}. More precisely, $C_V$ is a linear function of
$T$ for $T \ll T_F$, while it approaches $3/2Nk_B$ as $T
\rightarrow \infty$ \cite{Huang}. In order to illustrate the above
result we display in Fig.~7 the dependence of $C$ and the specific
heat $C_V$ on $T$ (in units $T_F$). This behavior can be explained
as follows: Let us consider two momentum distributions $n(k)$, one
for $T=0$ and another one for $T>0$. They are in essence
different, because for $T>0$ a certain number of fermions are
excited above the Fermi level $\epsilon_{F}$. Specifically,
fermions with energies of the order of $k_{B}T$ below
$\epsilon_{F}$ are excited to energies of the order of $k_{B}T$
above $\epsilon_{F}$. However, this holds only for fermions with
energies within about $k_{B}T$ of the Fermi level, while those
with other values of energy have no place to go --the states are
occupied \cite{Feynman}.

To sum up, thermal effects lead to a blurring of the Fermi surface
that means the distribution function $n(\epsilon)$ drops over the
range $\epsilon_F \pm k_BT$. As a consequence, in the case of low
temperature limit, only the momentum distribution (or the
occupation number) close to the Fermi surface is affected by $T$
and this leads to a linear dependence of $C$ on $T$. For $T\gg
T_F$, the momentum distribution is affected  for both low and high
values of $k$ in such a way so that the complexity $C$ tends to a
constant.

\section{Conclusions}
In the present work the recently proposed statistical measure of
complexity $C$ has been an issue under consideration. Information
theoretical measures (information entropy and disequilibrium) as
well as a statistical measure of complexity $C$ have been
calculated, in momentum space, for realistic Fermi systems, that
is nuclear matter, electron gas and liquid helium. The dependence
of the above measures on the strength of the correlations has been
analyzed and displayed. From the present analysis it should become
clear that the values of the above quantities  could be used as a
measure of the particle correlations of Fermi systems. We have
found that the complexity is an increasing function of the
correlations both for nuclear matter and liquid helium as expected
intuitively. However, in the case of electron gas, complexity
exhibits a different slope and in fact, the function of $C(r_s)$
has a maximum for a specific value of the correlation parameter
$r_s$. Additionally, we have found the interesting result that for
very strong correlations, where electron gas undergoes Wigner
crystallization, the complexity is independent of the correlations
and takes a constant value. In order to have a common measure of
the various information properties, we have displayed them as a
function of the discontinuity gap, $(1-Z_F)$. The most distinctive
feature of the above analysis, in the various systems,  is the
different behavior exhibited by $S_{\rm cor}$, $D_{\rm cor}$ and
$C$ as functions of $(1-Z_F)$. For the same values of $(1-Z_F)$
both the values and the trend of these quantities are different in
the various systems. Considering that, under certain
circumstances, $Z_F$ can be estimated experimentally, we obtain  a
first indication that information properties may be related with
experimental results.

Temperature also affects the momentum distribution of an ideal
Fermi gas and consequently all the related information properties.
We have found that for low values of the ratio $T/T_F$ the
complexity is a linear function of $T$. However, in the high
temperature limit (the well known classical Maxwell-Boltzmann
distribution), complexity is independent of $T$ and takes a fixed
value (exactly the same as in the case of Wigner crystallization).
Thus, regardless of the reason that causes the momentum
distribution to exhibit gaussian type dependence on the momenta
$k$, the value of the complexity is constant. Furthermore, we have
seen that the temperature dependence of $C$ is similar to that of
the specific heat $C_V$ in an ideal Fermi-gas \cite{Huang}, both
for low and high values of $T$. This is a second  indication that
one can relate the statistical measure of complexity $C$ with
experimental data (as the specific heat $C_{V}$). However, further
work is called for before we establish a clear connection between
information theoretical measures and experimental data.

As an epilogue, we would like to mention that physicists have been
carrying out research for decades, going beyond the mean-field
description of  quantum many-body systems, by taking into account
correlations among particles, a very important factor indeed
towards a better understanding of these systems. The effect of
correlations is connected intuitively with the concept of
complexity, in a qualitative, and somehow vague way. The present
work contributes to a quantification of complexity $C$ in
correlated Fermi systems, based on previous research for the
information entropy of the same systems \cite{Moustakidis05}. It
turns out that $C(T)$ and the specific heat $C_{V}(T)$ are similar
functions of the temperature $T$ as seen in Fig~7(a). In Fig.~7(b)
we plot $C_V(C)$. The dependence of $C_V$ on $C$ is approximately
linear. In fact there appear two regions of linear dependence with
a different slope, separated by a cross. The fitted expressions
are $C_V=-1.7353+1.7114C$ (region A) and $C_V=-6.3777+5.0783C$
(region B). In a sense, one may state that $C_{V}(T)$ can serve as
an index reflecting the expected increase of complexity as $T$
increases. Furthermore, it is seen that for temperatures $T\gg
T_{F}$ complexity reaches a plateau (saturation) i.e. it can no
longer increase. The textbook definition of the specific heat is
\emph{the measure of heat energy required to increase the
temperature of a unit of quantity of a substance by a unit
degree}. It is noted that here, we observe an empirical connection
of  such an "energy-like" quantity with complexity $C$ calculated
employing information entropy, which is not related directly to
the energy of the system in contrast to the traditional concept of
thermodynamic entropy.

\section*{Acknowledgments}
The authors would like to thank Dr. Paola Gori-Giorgi for
providing the data for the correlated electron gas and Dr. Saverio
Moroni for the data for the liquid $^3$He.


\newpage

\begin{figure}
\centering
\includegraphics[height=9.0cm,width=8.0cm]{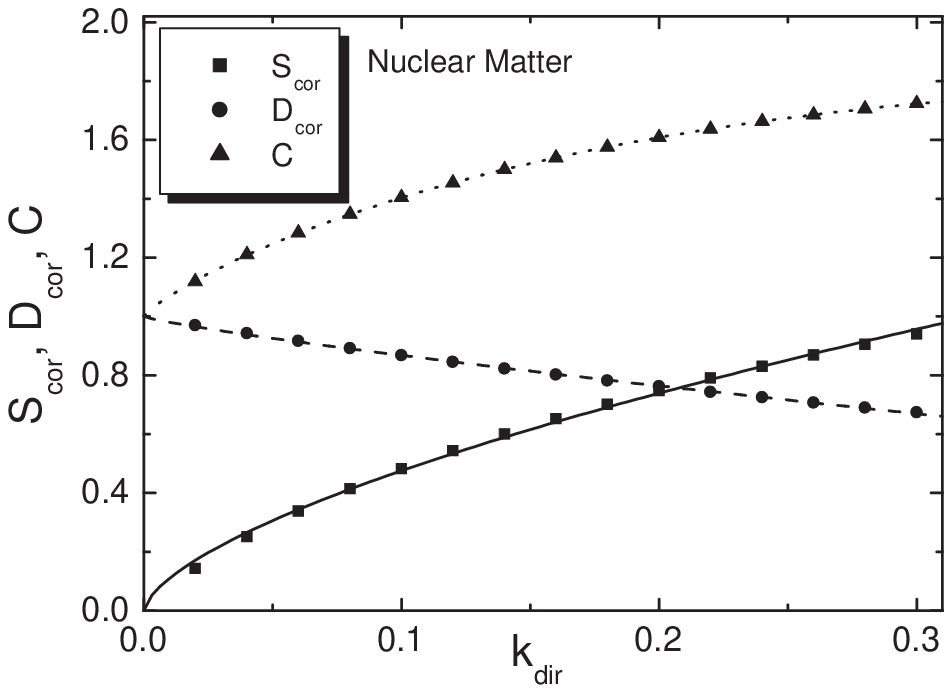}\
\caption{$S_{\rm cor}$, $D_{\rm cor}$ and $C$ of nuclear matter
versus the correlation parameter $k_{\rm dir}$. The lines
correspond to the expressions (\ref{Scor-fit-1}),
(\ref{Dcor-fit-1}), (\ref{Ccor-fit-1}), with the parameters
derived by the least squares fit method. } \label{}
\end{figure}
\begin{figure}
\centering
\includegraphics[height=8.0cm,width=8.0cm]{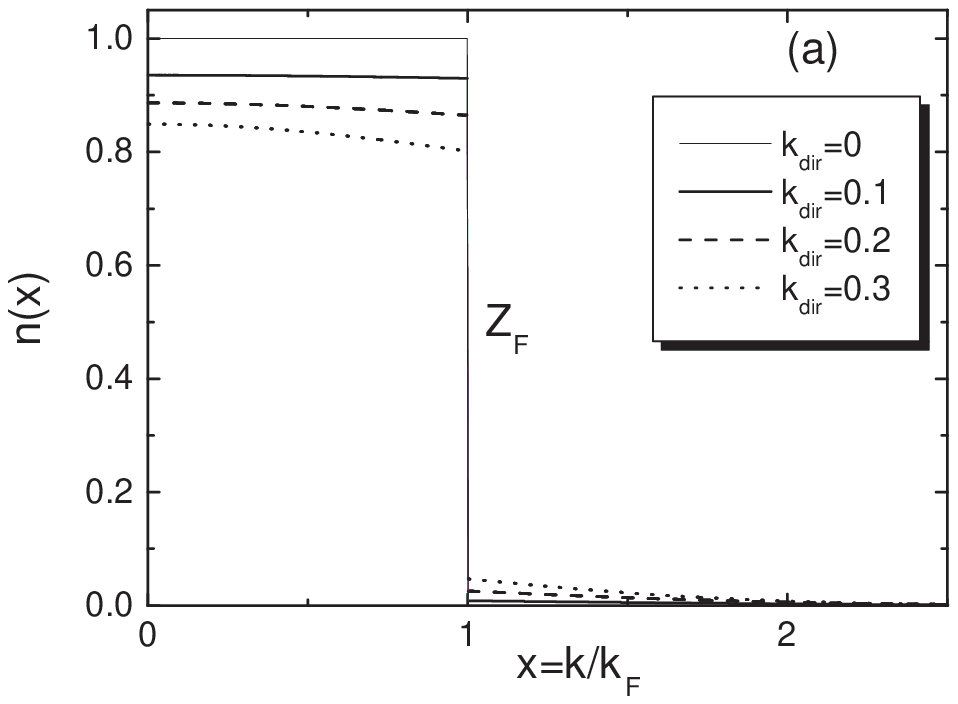}\
\includegraphics[height=8.0cm,width=8.0cm]{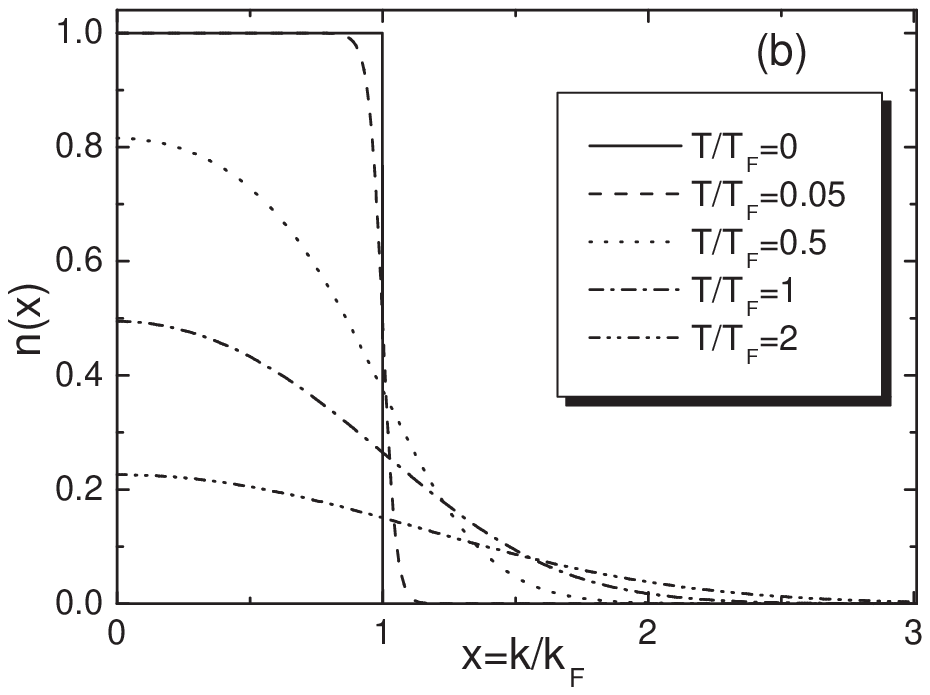}\
\caption{(a) The momentum distribution for correlated nuclear
matter versus $k/k_F$ for various values of the correlation
parameter $k_{\rm dir}$ (b) The momentum distribution of an ideal
electron gas versus $k/k_F$ for various values of the ratio
$T/T_F$.} \label{}
\end{figure}
\begin{figure}
\centering
\includegraphics[height=8.0cm,width=5.75cm]{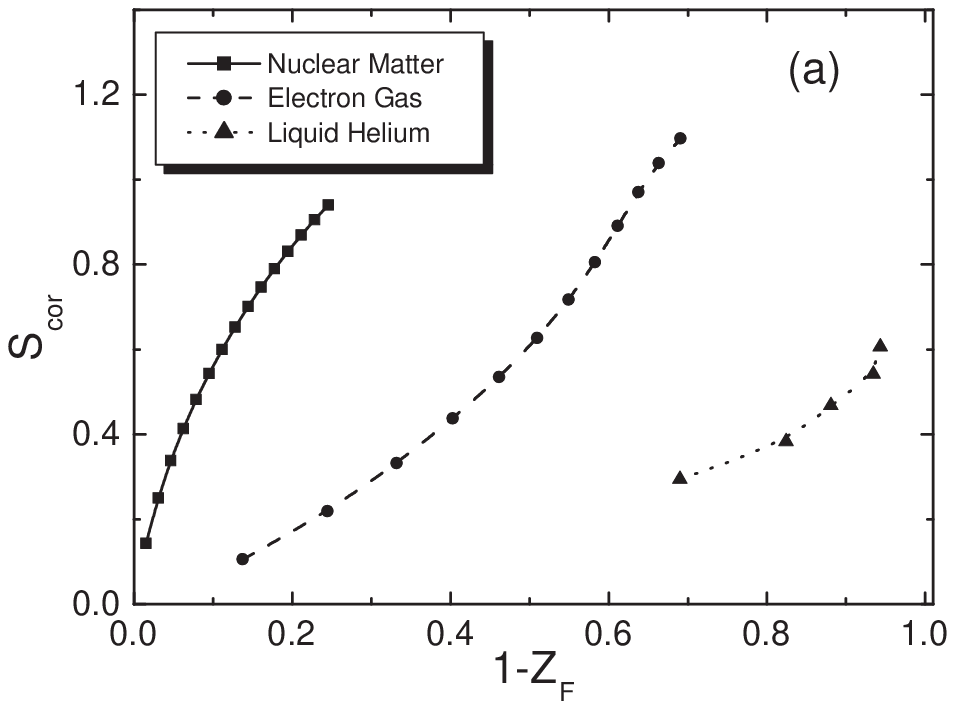}\
\includegraphics[height=8.0cm,width=5.75cm]{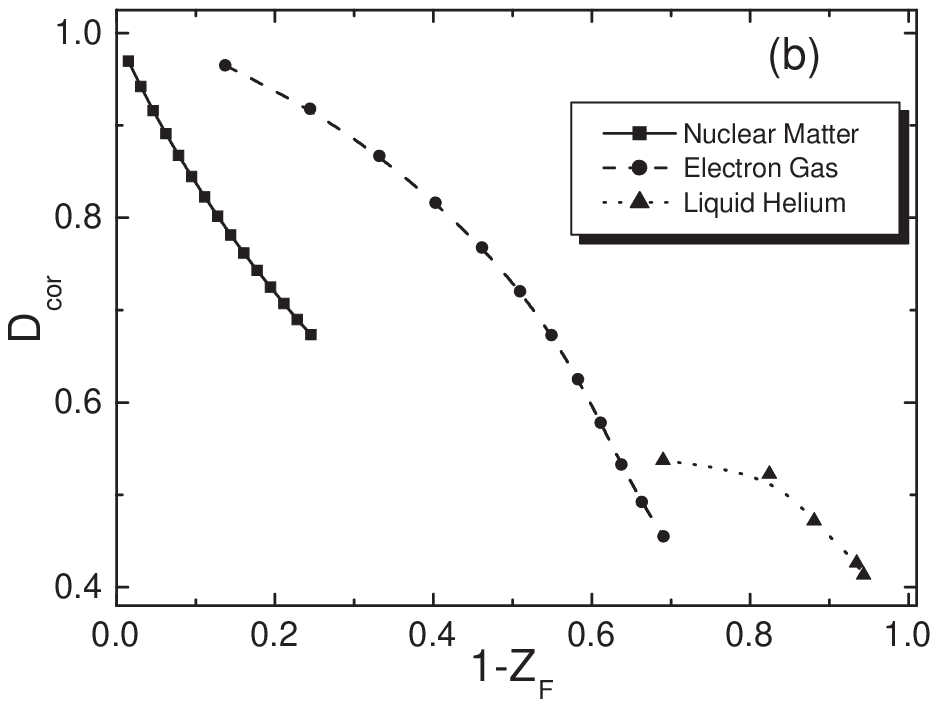}\
\includegraphics[height=8.0cm,width=5.75cm]{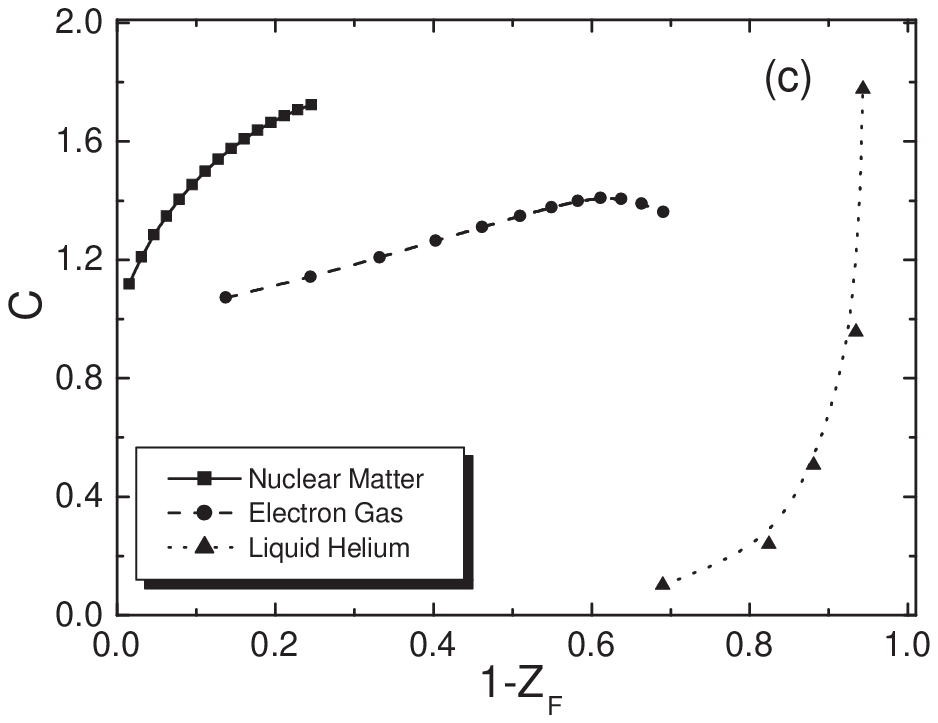}\
\caption{(a) $S_{\rm cor}$ (b) $D_{\rm cor}$ and (c) $C$ for
nuclear matter, electron gas and liquid helium versus the
discontinuity parameter $(1-Z_F)$. In the case of liquid helium,
the values of $S_{\rm cor}$ are divided by $10$ and the values of
$C$ by $100$. } \label{}
\end{figure}
\begin{figure}
\centering
\includegraphics[height=9.0cm,width=8.0cm]{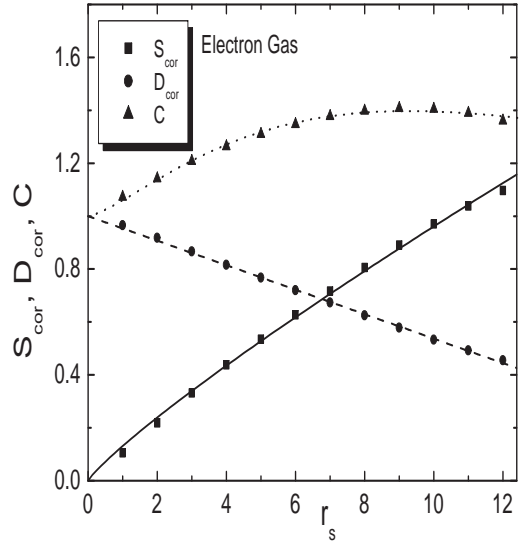}\
\caption{$S_{\rm cor}$, $D_{\rm cor}$ and $C$ of electron gas the
correlation parameter $r_s$. The lines correspond to the
expressions (\ref{Scor-ro-EG}), (\ref{Dcor-EG-1}),
(\ref{Ccor-EG-1}), with the parameters derived by the least
squares fit method. } \label{}
\end{figure}
\begin{figure}
 \vspace{3cm}
\centering
\includegraphics[height=9.0cm,width=8.0cm]{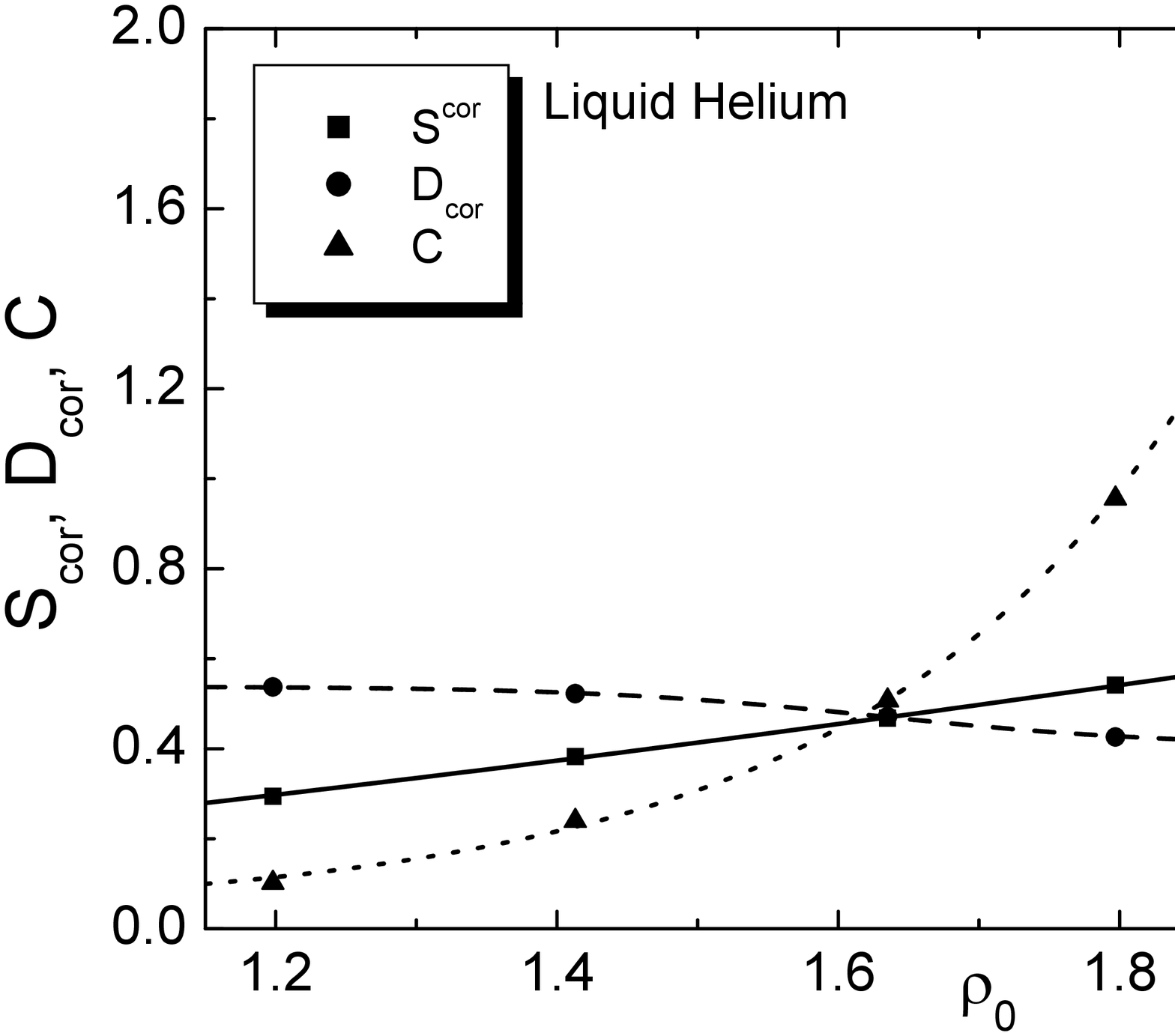}\
\caption{$S_{\rm cor}$, $D_{\rm cor}$ and $C$ of liquid helium
versus the correlation parameter $\rho_0$. The values of $S_{\rm
cor}$ are divided by $10$ and the values of $C$ by $100$. The
lines correspond to the expressions (\ref{Scor-rho-HL}),
(\ref{Dcor-LH-1}), (\ref{Ccor-LH-1}), with the parameters derived
by the least squares fit method.    } \label{}
\end{figure}
\begin{figure}
 \vspace{3cm}
\centering
\includegraphics[height=9.0cm,width=8.0cm]{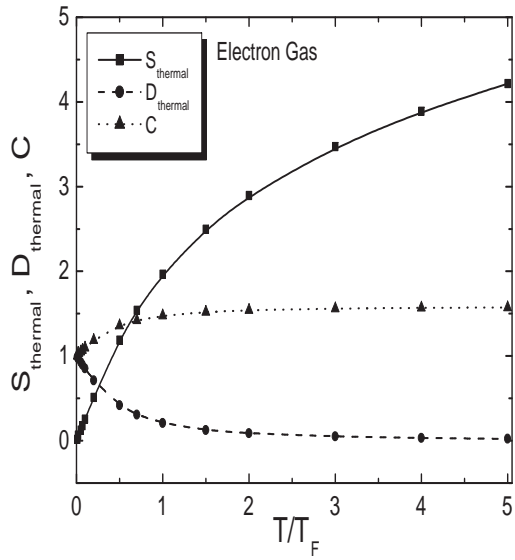}\
\caption{$S_{\rm thermal}$, $D_{\rm thermal}$ and $C$ of an ideal
electron gas versus the ratio $T/T_F$.} \label{}
\end{figure}
\begin{figure}
 \vspace{3cm}
\centering
\includegraphics[height=8.0cm,width=8.0cm]{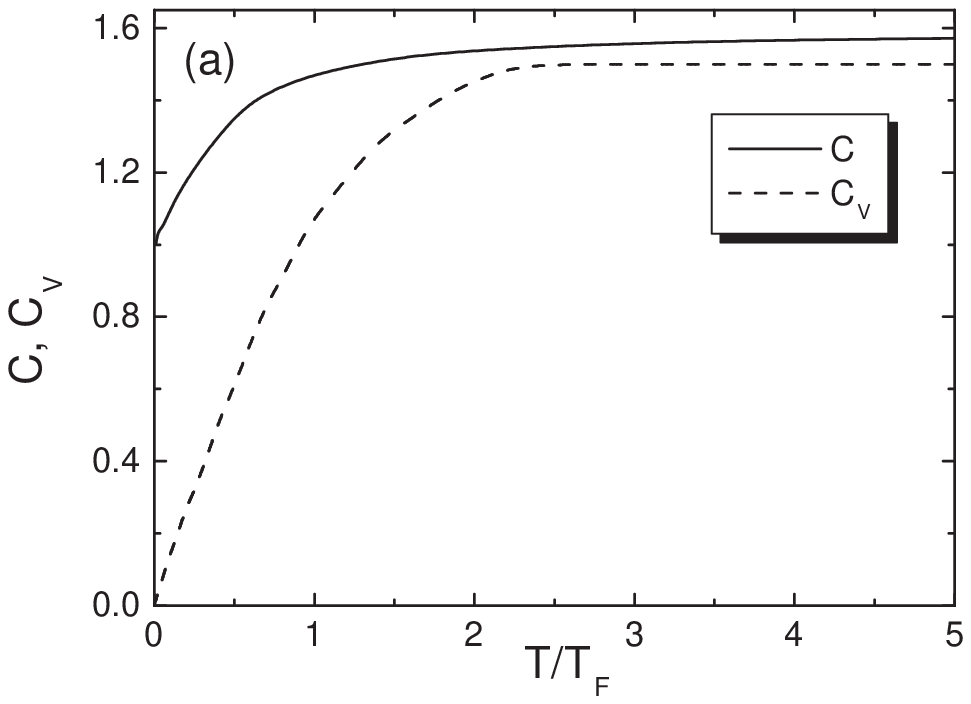}\
\includegraphics[height=8.0cm,width=8.0cm]{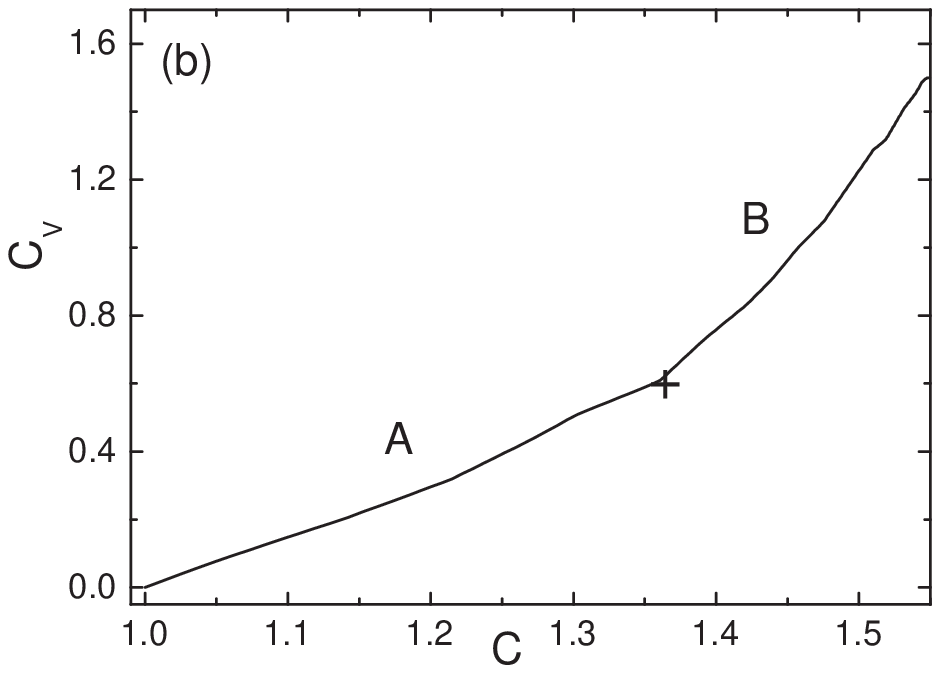}\
\caption{(a) The complexity $C$ and the specific heat $C_V$ (in
units $k_BN$) of an ideal electron gas versus the ratio $T/T_F$
(b) The specific heat $C_V$ versus complexity $C$.} \label{}
\end{figure}

\end{document}